\documentstyle[emulateapj]{article}
\def\etal{{et al.}\ }
\def\per{$^{-1}$}
\def\persq{$^{-2}$}
\def\hal{H$\alpha$}
\def\hbeta{H$\beta$}
\def\oiii{[\ion{O}{3}] $\lambda5007$}
\def\kms{km s$^{-1}$}
\def\persqcm{cm$^{-2}$}
\def\percucm{cm$^{-3}$}
\def\ncrit{$n_{crit}$}
\def\msun{$M{\footnotesize{_{\odot}}}$}
\def\percu{${^{-3}}$}
\def\hst{\emph{HST}}
\def\lam{$\lambda$}
\def\ebv{$E(B-V)$}

\begin{document}

\submitted{To appear in The Astronomical Journal}

\title{Polarized Narrow-Line Emission from the Nucleus of NGC 4258}

\author{Aaron J. Barth\altaffilmark{1},
Hien D. Tran\altaffilmark{2,3},
M. S. Brotherton\altaffilmark{2}
Alexei V. Filippenko\altaffilmark{4},
Luis C. Ho\altaffilmark{5},
Wil van Breugel\altaffilmark{2},
Robert Antonucci\altaffilmark{6},
and
Robert W. Goodrich\altaffilmark{7}}

\altaffiltext{1}{Harvard-Smithsonian Center for Astrophysics, 60
Garden St.,
Cambridge, MA 02138.}
\altaffiltext{2}{Institute of Geophysics and Planetary Physics, Lawrence
Livermore National Laboratory, 7000 East Avenue, P.O. Box 808, L413,
Livermore, CA 94550.}
\altaffiltext{3}{Department of Physics and Astronomy, Johns Hopkins
University, Baltimore, MD 21218.}
\altaffiltext{4}{Department of Astronomy, University of California,
Berkeley, CA 94720-3411.}
\altaffiltext{5}{Observatories of the Carnegie Institution of
Washington, 813 Santa Barbara St., Pasadena, CA 91101-1292}
\altaffiltext{6}{Department of Physics, University of California at
Santa Barbara, Santa Barbara, CA 93106.}
\altaffiltext{7}{W. M. Keck Observatory, 65-1120 Mamalahoa Highway,
Kamuela, HI 96743.}

\begin{abstract}

The detection of polarized continuum and line emission from the
nucleus of NGC 4258 by Wilkes \etal (1995) provides an intriguing
application of the unified model of Seyfert nuclei to a galaxy in
which there is known to be an edge-on, rotating disk of molecular gas
surrounding the nucleus.  Unlike most Seyfert nuclei, however, NGC
4258 has strongly polarized \emph{narrow} emission lines.  To further
investigate the origin of the polarized emission, we have obtained
spectropolarimetric observations of the NGC 4258 nucleus at the
Keck-II telescope.  The narrow-line polarizations range from 1.0\% for
[\ion{S}{2}] \lam6716 to 13.9\% for the [\ion{O}{2}]
$\lambda\lambda7319,7331$ blend, and the position angle of
polarization is oriented nearly parallel to the projected plane of the
masing disk. A correlation between critical density and degree of
polarization is detected for the forbidden lines, indicating that the
polarized emission arises from relatively dense ($n_e \gtrsim 10^4$
\percucm) gas.  An archival \emph{Hubble Space Telescope} narrow-band
[\ion{O}{3}] image shows that the narrow-line region has a compact,
nearly unresolved core, implying a FWHM size of $\lesssim2.5$ pc.  We
discuss the possibility that the polarized emission might arise from
the accretion disk itself and become polarized by scattering within
the disk atmosphere.  A more likely scenario is an obscuring torus or
strongly warped disk surrounding the inner portion of a narrow-line
region which is strongly stratified in density.  The compact size of
the narrow-line region implies that the obscuring structure must be
smaller than $\sim2.5$ pc in diameter.

\end{abstract}

\keywords{galaxies: active -- galaxies: individual (NGC 4258) ---
galaxies: nuclei -- galaxies: Seyfert --- polarization}

\section{Introduction}

The low-luminosity Seyfert galaxy NGC 4258 provides some of the most
convincing evidence for a link between supermassive black holes and
nuclear activity.  Its nucleus exhibits hard X-ray emission
(\cite{mak94}), broad \hal\ emission (\cite{fs85}; \cite{hfsp97}), and
jets which have been observed on subparsec scales in radio continuum
emission (\cite{her97a}).  The kiloparsec-scale ``anomalous arms''
observed in optical emission lines, X-rays, and radio emission
(\cite{cc61}; \cite{cmv95} and references therein) are thought to be
the outer extensions of the nuclear jet (but see \cite{cd96} for an
alternative interpretation).  Based on these observations, it is clear
that NGC 4258 contains a genuine low-luminosity active nucleus.  The
discovery of a rotating molecular disk emitting H$_2$O maser lines
provides a unique means to trace out the nuclear rotation curve in the
inner parsec.  Miyoshi \etal (1995) have shown that the disk rotation
is Keplerian and implies a central mass of $3.6 \times 10^7$ \msun\
within a radius of 0.13 pc (for a distance of 6.4 Mpc).  Alternatives
to a supermassive black hole are implausible, because the derived
central mass density is so large that a hypothetical cluster of dark
objects would have a lifetime of at most $\sim5 \times 10^8$ yr
against evaporation and/or collisions (\cite{maoz98}).  This
combination of nuclear activity, a well-determined central mass, and a
rotating circumnuclear disk makes NGC 4258 a particularly interesting
laboratory in which to study many aspects of the AGN phenomenon.

NGC 4258 is a natural target for spectropolarimetric investigation,
because the nuclear disk is nearly edge-on.  Wilkes \etal (1995;
hereafter W95) discovered that the emission lines and continuum in NGC
4258 are polarized, with the angle of polarization nearly parallel to
the disk plane (P.A. = 86\arcdeg; Miyoshi \etal 1995).  This
coincident orientation was interpreted as evidence that the
polarization is due to scattering by electrons or dust grains in
clouds located above the plane of an optically thick torus.  The
discovery of polarized emission lines in NGC 4258 has important
implications for unified model of AGNs, as it provides a direct link
between polarized nuclear emission and the presence of an edge-on,
parsec-scale molecular disk.  It also suggests that the unified scheme
can apply to even the lowest luminosity Seyfert nuclei.  One
interesting aspect of the W95 results is the unusually high
polarization of the narrow emission lines (4.6\% for [\ion{O}{1}]
\lam6300).  In typical Seyfert 2 nuclei the narrow lines are
unpolarized, or are weakly polarized by transmission through dust in
the host galaxy (\cite{goo92}).  In this paper, we focus on the
polarization mechanism of the narrow emission lines, using Keck
spectropolarimetry data and archival \emph{Hubble Space Telescope
(HST)} images.  In the following discussion, we assume a distance of
7.3 Mpc to NGC 4258, based on recent observations of proper motions in
the maser disk (\cite{her97b}); at this distance, 1\arcsec\
corresponds to 35 pc, and the central mass measured by Miyoshi \etal
(1995) scales to $4.1 \times 10^7$ \msun.

\section{Observations and Results}

\subsection{Spectropolarimetry}

The observations were obtained on 1997 April 10 UT at the Keck-II
telescope using the LRIS spectropolarimeter (\cite{oke95};
\cite{coh97}).  The seeing was 0\farcs6--0\farcs7, and conditions were
photometric.  We used a 300 grooves mm\per\ grating blazed at 5000
\AA\ and a 1\arcsec\ slit, yielding a spectral range of 3800--8700
\AA, and the spatial scale was 0\farcs215 pixel\per. The spectral
resolution, as measured from a comparison lamp exposure, varied from
10 \AA\ at the center of the spectrum to $\sim13$ \AA\ at the red and
blue ends.  The exposure time was 5 minutes for each of the four
waveplate positions.  The spectrograph slit was oriented east-west
(P.A. = 90\arcdeg) during the observations, while the parallactic
angle was 50\arcdeg; this offset affects the shape of the polarized
continuum at the blue end of the spectrum, as we discuss below.  No
order-blocking filter was used, and there may be some second-order
light longward of 7600 \AA, but none of our conclusions are affected
by this contamination.

The data were reduced with the VISTA software package (\cite{tls84})
according to the methods described by Cohen \etal (1997) and Miller
\etal (1988), using the polarized standard star HD 155528
(\cite{ct90}) to calibrate the position angle of polarization.
Spectral extractions were performed using a width of 3\farcs4 along
the slit, and the results are shown in Figure \ref{4258montage}.  A
1\farcs1-wide extraction was also performed, in order to better
isolate the nuclear emission, but the continuum polarization in the
narrow extraction is more severely affected by the slit misalignment.
The emission-line results for the two extractions are substantially
similar, with most emission-line polarizations agreeing to within
$1\sigma$, and all results in this paper refer to the 3\farcs4-wide
extraction.

To measure accurate line polarizations, starlight was subtracted from
the total flux spectrum using the methods described by Ho \etal (1993)
and Tran (1995), with a spectrum of the M31 nucleus used as a
template.  We were able to achieve a satisfactory continuum
subtraction without including any nonstellar contribution, and we
estimate that a nonstellar continuum source can contribute at most 5\%
of the flux in our aperture (provided that the stellar populations in
the NGC 4258 and M31 nuclei have similar metallicities).  All
polarization measurements were carried out on the Stokes parameter
($q$ and $u$) spectra, with the results converted to degree and
position angle of polarization ($p$ and $\theta$) only as the final
step.  The polarizations of unblended lines were measured by direct
integration of the fluxes and Stokes parameters using the methods
outlined by Miller \etal (1988).  Table 1 lists the individual emission-line
polarizations.

The polarizations of the individual components of the
\hal+[\ion{N}{2}] and [\ion{S}{2}] blends were measured by fitting the
emission profiles in the total flux ($f$) spectrum and in the $f
\times q$ and $f \times u$ spectra, using the SPECFIT package within
IRAF.  Our observations do not have sufficient spectral resolution to
determine whether there is a broad component of \hal\ in total flux; a
good fit to the \hal+[\ion{N}{2}] profile can be achieved either with
or without the inclusion of a broad \hal\ component.  The higher
resolution observations of Ho \etal (1997a) demonstrate that the total
flux spectrum contains a broad \hal\ component of FWHM $\approx 1700$
\kms\ which is not present in the profiles of the narrow forbidden
lines, but the Ho \etal observations were obtained in 1984 and the
broad-line flux could be variable\footnote{We note that W95 do not
discuss whether a broad component of \hal\ may be present in polarized
flux.  Their 400 \kms\ spectral resolution is somewhat higher than
that of our observations but still not sufficient to permit the
unambiguous detection of a 1700 \kms\ broad component of \hal\ by
fitting the \hal+[N II] blend.}.  Therefore, we measured the \hal\ and
[\ion{N}{2}] component polarizations by fitting multi-Gaussian models
both with and without a broad \hal\ component.  Since our data do not
clearly demonstrate the existence of the broad \hal\ component, we
take as our ``default'' results the line polarizations measured by
fitting the blend without a broad \hal\ component; these results are
listed in Table 1.

In the fits which do include a broad \hal\ component, the width of
broad \hal\ is not well constrained, so the broad component was fixed
to have FWHM = 1700 \kms\ to match the results of Ho \etal (1997a).
For these fits, the resulting polarizations are $p = 12.2\% \pm 1.1\%$
for broad \hal, $p = 2.7\% \pm 0.6\%$ for narrow \hal, and $p = 3.4\%
\pm 0.6\%$ for [\ion{N}{2}] \lam6584.  The quoted uncertainties
represent the formal errors of the fits, while the actual
uncertainties must be significantly larger because the data do not
show whether the broad \hal\ component is in fact present.

The Keck data confirm the general results found by W95: the narrow
emission lines are polarized with position angles (PAs) of typically
$89-90\arcdeg$, close to the projected 86\arcdeg\ PA of the masing
disk.  We find somewhat higher line polarizations than those measured
by W95, who used a $3\arcsec\times7\arcsec$ effective aperture, most
likely because our smaller $1\arcsec\times3\farcs4$ aperture admits
less unpolarized emission-line light from the outer regions of the
narrow-line region (NLR).  Similarly, the continuum polarization of
$0.35\%$ in the $V$-band region is greater than that measured by W95,
who found $p=0.23\%$ over 5100--6500 \AA.  Most of this continuum
polarization must be intrinsic to NGC 4258 rather than due to
transmission through Galactic dust, since a Galactic reddening of
$E(B-V) = 0.016$ mag (\cite{sfd98}) should result in a polarization of
at most 0.14\% (\cite{smf75}).  Furthermore, the similar position
angles of the line and continuum polarization suggest that the
continuum polarization is not dominated by interstellar dust
transmission.

The linewidths listed in Table 1 were measured by fitting Gaussian
profiles to each emission feature, and subtracting the instrumental
resolution in quadrature from the measured widths.  Figure
\ref{4258fluxcomp} compares the emission-line profiles in total flux
and Stokes flux.  As found by W95, the emission lines are broader in
polarized light than in the total flux spectrum.  This is most clearly
visible for \oiii, which has FWHM = 530 \kms\ in total flux and 1010
\kms\ in Stokes flux.  The \hal+[\ion{N}{2}] blend has nearly the same
FWHM in total and Stokes flux, but a broader base appears in Stokes
flux, particularly on the red side of the profile, and the narrow
\hal\ peak is broader in Stokes flux.

Figure \ref{4258fluxcomp} also shows that the [\ion{O}{3}] \lam5007
line is blueshifted in Stokes flux relative to its profile in total
light.  In the observed wavelength frame, the \lam5007 line has
centroid $5011.0 \pm 0.6$ \AA\ in total flux and $5008.9 \pm 0.4$ \AA\
in Stokes flux, for a shift of 126 \kms\ with a significance level of
$2.9\sigma$.  The other lines are too noisy in Stokes flux to
determine whether there is a significant velocity shift relative to
the profiles in total flux.

One remarkable feature of the Stokes flux spectrum is the weakness of
polarized [\ion{S}{2}] emission relative to the other emission
features (as noted by W95), and the fact that the [\ion{S}{2}]
$\lambda6716/6731$ flux ratio appears to be greater in total light
than in Stokes flux.  Such a change in the relative strengths of two
closely spaced lines cannot be a result of the interaction with the
scattering medium; the more likely cause is that the scattering medium
``sees'' a different [\ion{S}{2}] ratio, and therefore a different
density, than what is observed in the total flux spectrum.  In total
flux, $I(\lambda6716)/I(\lambda6731) = 0.94$, corresponding to a
density of $n_e = 750$ cm\percu\ at $T_e = 10^4$ K.  The [\ion{S}{2}]
ratio in polarized light is $0.31 \pm 0.20$, formally lower than (but
consistent with) the theoretical minimum value of 0.44 for the
high-density limit, and yielding a density of $n_e \gtrsim 10^4$
cm\percu\ (\cite{cka86}).

The other forbidden lines exhibit similar behavior.  Figure
\ref{4258pvsnc} shows the degree of polarization of the forbidden
lines as a function of the critical density for collisional
de-excitation (\ncrit) of each line, and also as a function of the
ionization potential (IP) of the ionization state giving rise to each
line.  A trend is apparent: although there is considerable scatter,
lines of greater \ncrit\ tend to be more highly polarized than lines
of lower \ncrit, and the trend appears to apply at least in a rough
sense over nearly four orders of magnitude in \ncrit.  The
significance of the correlation between $p$ and \ncrit\ depends on
which value for $p$([\ion{N}{2}] \lam 6584) is used: the value of
7.7\% derived from the 3-Gaussian fit to the \hal+[\ion{N}{2}] blend,
or the lower value of 3.4\% derived from the 4-Gaussian fit with a
broad \hal\ component.

For the $p$-\ncrit\ relation using the value of $p$([\ion{N}{2}])
derived from the fit without a broad \hal\ component, we find a
Spearman rank correlation coefficient of $r_s=0.68$, which would arise
by chance between unrelated variables only 9.4\% of the time (a
two-tailed probability).  The correlation coefficient is similarly
high if lines with nearly equal values of $p$ are given identical
rank.  A marginally significant correlation is apparent between
polarization and IP ($r_s=0.59$, 16\% probability of arising by
chance).  Using the lower value of $p$([\ion{N}{2}]) = 3.4\% from the
4-Gaussian fit, however, the correlation between $p$ and \ncrit\
improves to $r = 0.89$, with only an 0.7\% probability of arising by
chance.  Since the higher-resolution spectra of Ho \etal (1997a)
indicate that a significant broad-line component of \hal\ was present
in 1984, the lower value of $p$([\ion{N}{2}]) = 3.4\% from the
4-Gaussian fit may be preferable to the ``default'' higher value of
$p$ = 7.7\% obtained from the 3-Gaussian fit.

Similar correlations, but between linewidth and \ncrit, have been
discovered in Seyferts and LINERs (e.g, \cite{paf81}; \cite{fh84};
\cite{f85}), although NGC 4258 is the first Seyfert known to show a
relationship between $p$ and \ncrit.  The Keck data do not have
sufficient spectral resolution to properly evaluate a possible
correlation between linewidth and \ncrit, but such a correlation may
be present, as the high-\ncrit\ lines of [\ion{O}{1}] and [\ion{O}{3}]
are broader than the low-\ncrit\ lines of [\ion{N}{2}] and
[\ion{S}{2}].  The large FWHM of 870 \kms\ for [\ion{O}{2}] \lam7325
is partly due to the fact that it is a blend of components with rest
wavelengths at 7319 and 7331 \AA, with possible [\ion{Ca}{2}] emission
at 7324 \AA\ as well.

\subsection{Archival HST Images}
\label{sectionhst}

To search for morphological evidence for an obscuring torus, we have
examined an archival \hst\ WFPC2/PC narrow-band image of the nucleus
of NGC 4258 (Figure \ref{4258oiii}).  The 2300 s exposure was taken on
1995 March 16 through the narrow-band F502N filter, which isolates the
[\ion{O}{3}] \lam5007 line.  Continuum emission was subtracted using
an image of the same field taken in the F547M ($V$-band) filter.  The
F547M image was scaled so that the maximum amount of continuum
emission was removed without leaving negative ``holes'' around the
nucleus or at the positions of other stars in the field.

If the nuclear disk flares out to a thick obscuring torus extending
out to radii of a few parsecs, such a structure might be apparent in
the [\ion{O}{3}] image as a dark band across the nucleus.  Similarly,
an ionization cone extending north-south would be another possible
consequence of a thick torus.  The \emph{HST} image shows that the
[\ion{O}{3}]-emitting region is strongly peaked at the nucleus, with
some extended emission surrounding the nuclear peak.  The central
spike of [\ion{O}{3}] emission is only marginally more extended than a
synthetic F502N point-spread function generated using the TinyTim
package (\cite{kh97}), implying a FWHM size of $\lesssim2.5$ pc.  The
F547M continuum image, on the other hand, does not show evidence for a
strong nuclear point source.  This resolved continuum morphology is
typical of Seyfert 2 nuclei (\cite{nel96}; \cite{mgt98}).

Some interesting structure is seen in the extended [\ion{O}{3}]
emission on larger scales.  The off-nuclear emission is concentrated
in a faint region extending 2\arcsec\ south of the nucleus and in a
bright arc extending from $\sim2\arcsec-5\arcsec$ north of the
nucleus, with prominent knots or filaments along the northwest portion
of the arc.  Since the extended emission is concentrated in patches
located above and below the plane of the masing disk, these regions
may represent ionization cones illuminated by anisotropic ionizing
radiation from the AGN, or regions in which the nuclear jet interacts
with interstellar clouds (\cite{ftk96}).

The \hst\ images also allow us to make a quantitative assessment of
whether the 40\arcdeg\ offset between the slit angle and the
parallactic angle may have had a negative impact on our data.  At
airmass 1.26, atmospheric dispersion will displace the emission from a
point source by $\sim0\farcs25$ at 5000 \AA, and by $\sim0\farcs61$ at
4000 \AA, relative to the position of the source at 6500 \AA\
(\cite{fil82}; recalculated for conditions appropriate to Mauna Kea).
Since the LRIS guide camera views the red portion of the spectrum, the
blue image of the nucleus would have appeared off-center in the slit.

To assess the effects of this misalignment, we performed aperture
photometry on the \hst\ images using a $1\arcsec\times3\farcs4$
rectangular aperture.  The results that were obtained with the
aperture centered on the nucleus were then compared with results
obtained with the aperture centered off the nucleus at a position
determined by the atmospheric dispersion.  For the simulations, the
``nuclear'' region was defined as that portion of the image within a
radius of $r=5$ PC pixels (0\farcs23) surrounding the nucleus, which
includes nearly all the light from the central [\ion{O}{3}] emission
spike.  The images were convolved with a Gaussian of FWHM = 0\farcs7
prior to measurement, to match the seeing during the Keck run.

The simulations indicate that the observed fraction of [\ion{O}{3}]
\lam5007 light coming from the nuclear region is 0.54, compared with a
nuclear fraction of 0.56 that would have been obtained had the
spectrograph slit been properly oriented at the parallactic angle.
Assuming that the polarized emission originates from the central
spike, this result implies that the measured polarization of
[\ion{O}{3}] is 0.96 times the polarization that would have been
measured with the slit at the parallactic angle.  Similarly, the
measured nuclear fraction of the 5000 \AA\ continuum is 0.95 times the
value that would have been obtained at the parallactic angle.  Thus,
the region of the spectrum which includes most of the important
emission lines is only slightly affected by the spectrograph slit
misalignment.  Blueward of 5000 \AA, however, the measurements are
more severely affected.  At 4000 \AA, the simulations indicate that
the observed nuclear fraction is only 0.77 times the value that would
have been measured with the slit at the parallactic angle.  Thus, the
$p$ and Stokes flux spectra shown in Figure \ref{4258montage} should
be revised upward by a factor of approximately 1.3 at the extreme blue
end of the observed wavelength range, provided that the polarized
emission source is as compact as we have assumed.

\section{Discussion}

NGC 4258 appears analogous in some respects to higher-luminosity
Seyfert 2 nuclei, but its narrow-line polarization properties are
unlike those of most Seyfert 2 nuclei having hidden broad-line
regions\footnote{One other example of a Seyfert 2 galaxy having
broadened forbidden lines in its polarized flux spectrum is IRAS
20210+1121 (\cite{you96}).  Young \etal propose that a portion of the
NLR in this galaxy is obscured and that the narrow-line polarization
is the result of scattering outside the obscuring region.}.  The X-ray
continuum is heavily obscured ($N_H \approx 1.5 \times 10^{23}$
\persqcm; \cite{mak94}), but it is not known whether this same
obscuring material is distributed on large enough scales to cover a
portion of the NLR as well.  Since the narrow-line properties of NGC
4258 are so unusual, and there is no clear evidence for polarized
broad emission lines, it is worthwhile to consider whether the
obscuring torus model applies to this galaxy, or whether an entirely
different mechanism could be responsible for the emission-line
polarization.

Polarization by transmission through foreground dust is ruled out by
the broadening of the line profiles seen in polarized light, and by
the high line polarizations; a foreground polarization of 10\% would
require a high reddening of \ebv $> 1$ mag.  The narrow-line
\hal/\hbeta\ intensity ratio is 3.94 (\cite{hfs97}), which corresponds
to a reddening of \ebv\ = 0.24 mag for an intrinsic \hal/\hbeta\ ratio
of 3.1, the value appropriate for gas photoionized by an AGN continuum
(\cite{fn83}; \cite{hs83}).  Thus, the only viable polarization
mechanism for the emission lines is scattering, either by dust
particles or by electrons.  In the following discussion, we consider
the possibility that our view of the accretion disk is unobscured, and
that emission lines from the disk surface are polarized by scattering
within the disk atmosphere.  We conclude that a more likely
explanation for the narrow-line polarization is obscuration of the
inner NLR by a thick torus or highly warped disk, combined with
scattering above the torus midplane.  In this case, our observations
allow some constraints to be set on the properties of the torus and
the scattering region.

\subsection{Scattering in the accretion disk atmosphere}

We first consider the possibility that the nuclear disk is neither
sufficiently warped nor sufficiently thickened at large radii to
obscure a substantial part of the NLR.  In this case, we would have a
direct but oblique view of the surface of the accretion disk.  Because
of the warped shape of the disk, portions of the disk face are exposed
to X-ray emission from the central engine.  As discussed by Neufeld \&
Maloney (1995) and Herrnstein \etal (1996), the X-ray irradiation
causes dissociation of molecules in the surface layers of the disk,
creating a layer of warm ($\sim8000$ K) partially ionized gas
surrounded by a fully ionized envelope.  Since the X-ray flux drops
off more slowly as a function of $r$ than the disk midplane pressure,
the ionization parameter actually increases with radius (\cite{nm95}).
The ionized outer regions of the disk should emit a narrow-line
spectrum typical of that of Seyfert or LINER nuclei, but with an
enhancement in high-\ncrit\ lines due to the high densities in the
disk.

Radiation emerging from the surface of an accretion disk is expected
to be polarized by electron scattering in the ionized atmosphere of
the disk.  For an electron-scattering atmosphere with infinite optical
depth, the numerical results of Chandrasekhar (1960) show that the
expected polarization is $p=11.7\%$ at $i=90\arcdeg$, dropping to
9.0\% at $i=87\arcdeg$ and 7.5\% at $i=84\arcdeg$, with the
polarization angle oriented parallel to the projected surface of the
disk.  Since the observed emission-line polarizations are oriented
parallel to the disk plane, it is worthwhile to consider whether this
type of polarization due to disk emission could be occurring in NGC
4258.  In this scenario, the correlation between $p$ and \ncrit\ would
be ascribed to the tendency for lines of high \ncrit\ to be emitted
preferentially by dense gas in the disk, and polarized in the disk
atmosphere, while lines of low \ncrit\ would have larger, unpolarized
contributions from NLR gas surrounding the disk.  Because of the
nearly edge-on orientation of the disk, the X-ray obscuring column
could be provided by a warp in the masing disk rather than by a
geometrically thick torus.

However, it is unlikely that the NGC 4258 disk could have the
structure required to generate the observed high polarizations in this
manner.  The detailed discussion of accretion disk polarization by
Chen \etal (1997) shows that two conditions are required in order to
generate polarizations of up to 11.7\% oriented parallel to the disk
plane.  The first is that the emission-line source must be located
below the scattering atmosphere.  An emission-line source mixed
cospatially with the scattering atmosphere would generate polarization
perpendicular to the disk plane, unlike what is observed in NGC 4258.
Second, the Chandrasekhar results are valid for scattering optical
depths exceeding $\tau \approx 4$.  Lower optical depths would lead to
polarizations much less than 11.7\%, or polarizations oriented
perpendicular to the disk axis.  However, an electron-scattering
atmosphere with a high optical depth would prevent ionizing photons
from reaching a line-emitting layer located below the scattering
atmosphere, and some other mechanism such as shock heating in the disk
would be required to generate the emission lines.  Thus, external
illumination of an accretion disk would most likely lead to
polarizations of $<10\%$, possibly oriented perpendicular to the disk
plane, unlike what is observed in NGC 4258.

Emission from a disk might be expected to produce double-peaked
emission profiles, which have not been observed in NGC 4258, but
electron scattering in the disk atmosphere could mask this signature
by broadening the profiles.  Low-velocity emission from narrow-line
gas surrounding the disk at larger radii would also tend to fill in
the centers of the profiles. 

Even if the accretion disk were somehow able to produce
Chandrasekhar-type polarization, the observed emission-line
polarizations in NGC 4258 are too high to be consistent with this
model because the disk is not quite edge on, and because there is some
extended emission from the surrounding NLR, which we assume to be
unpolarized.  For [\ion{O}{2}] \lam7325, the polarization of $13.9\%
\pm 1.4\%$ exceeds the theoretical maximum of 11.7\% for the
disk-emisison model, although only by $1.6\sigma$.  The best-fitting
warped disk models of Herrnstein \etal (1996) indicate that the masing
portion of the disk has $i = 82-84.5\arcdeg$, for which the expected
polarization of disk emission is only $6.5-7.5\%$.  The [\ion{O}{3}]
line polarization falls within this range, but the \hst\ image shows
that nearly half of the [\ion{O}{3}] flux in the Keck aperture does
not come from the unresolved central source.  Thus, even under the
assumption that the unresolved [\ion{O}{3}] source is a disk with
intrinsic polarization of 7\%, the off-nuclear emission would dilute
the observed [\ion{O}{3}] polarization to a level of $p \approx
3.5\%.$

Therefore, despite the intriguing possibility that the emission-line
polarizations might be generated within the accretion disk, we
consider it unlikely that this scenario applies to NGC 4258.  This
leaves the standard obscuring torus model as the most likely
polarization mechanism.

\subsection{Scattering outside an obscuring torus}

In Seyfert 2 nuclei with polarized broad-line emission, such as NGC
1068 (\cite{am85}), the polarization is generally thought to result
from scattering by dust or electrons located in the opening cone of an
optically and geometrically thick torus.  W95 argued in favor of this
model for NGC 4258 in light of the fact that the continuum and
emission lines are polarized nearly parallel to the disk plane.  The
high narrow-line polarizations (7.2\% for [\ion{O}{3}] and 13.9\% for
[\ion{O}{2}] \lam7325) imply that a substantial fraction of the NLR
must be obscured from direct view.  Therefore, for this model to apply
to the narrow emission lines in NGC 4258, the disk must be either
opaque and highly warped, or it must be surrounded by a thick torus,
at radii larger than the size of the masing region.

In the context of the obscuring torus interpretation, the observed
dependence of $p$ on \ncrit\ can be interpreted as resulting from the
combined effects of density stratification of the NLR gas and
obscuration of the NLR by a thick torus or strongly warped disk.
Lines of higher \ncrit\ are preferentially emitted from denser
portions of the NLR lying closer to the central source (possibly by
the disk itself), and hidden from direct view by the obscuring torus,
and lines of lower \ncrit\ would be emitted by more diffuse gas at
larger radii, and less strongly affected by the obscuring torus and
scattering medium.  A similar phenomenon has been proposed to explain
observations of polarized \oiii\ emission in radio galaxies
(\cite{dsa97}).

Some constraints on the size of the obscuring torus can be derived
from the compact [\ion{O}{3}] morphology.  Since the core of the
[\ion{O}{3}] emission is nearly unresolved, the angular size of the
obscuring torus cannot be much larger than the size of the
[\ion{O}{3}] core in the \hst\ image, or 2.5 pc.  A larger torus would
most likely be visible as a dark band across the nucleus.  The
possibility that the obscuring material is in the form of a highly
warped thin disk rather than a thick torus should also be considered,
but the same constraints on the size of the obscuring material would
apply.  As an alternative to the warped disk models, Kartje \etal
(1999) have proposed that the masing clouds are lifted above the disk
surface in a hydromagnetically driven wind.  The dusty outer portions
of a disk-driven wind could constitute the obscuring torus (e.g.,
\cite{kk94}), and in this case the torus size would be naturally
related to the size of the masing disk.  

The broadening of [\ion{O}{3}] in polarized light may be a result of
interaction with the scattering medium, but to some extent it must
reflect the greater velocities of clouds which are located closer to
the nucleus and obscured by the torus.  For electron scattering, it is
possible to determine an upper limit to the scattering medium
temperature from the amount of line broadening.  Using the quantity
$\Delta u_{th}$ to represent the difference in quadrature between the
line FWHM in polarized light and in total light, the electron
temperature in the scattering medium is limited by $T_e \leq m_e
\Delta u_{th}^2 / (16 k \ln 2)$ (\cite{mgm91}).  From the [\ion{O}{3}]
linewidths, we find $T_e \leq 4400$ K.  As a comparison, in NGC 1068
the scattering medium is found to have an electron temperature of a
few $\times 10^5$ K (\cite{mgm91}).  The unusually small degree of
line broadening in NGC 4258 could be an indication that dust
scattering, rather than electron scattering, is the dominant
polarization mechanism.

Using FWHM/2 as an estimate of the Keplerian velocity, the
[\ion{O}{3}] linewidth of 530 \kms\ in total light corresponds to a
typical radius of 1.6 pc, which is six times larger than the outer
radius of the masing portion of the disk.  If the [\ion{O}{3}] line in
polarized light is not significantly broadened by scattering, then the
linewidth of 1010 \kms\ corresponds to a typical radius of $r \approx
0.7$ pc, small enough to fit within a torus of diameter 2.5 pc.  On
the other hand, if the torus is smaller than $\sim1.5$ pc in diameter
or if the emission lines are broadened by scattering, this would imply
that the high-density portion of the NLR, from which the polarized
emission originates, is larger than the size of the torus; this would
pose difficulties for the obscuring torus interpretation.

W95 estimated a size of $\sim35$ pc for the continuum scattering
region, based on their imaging polarimetry observations.  However, the
compact morphology of the [\ion{O}{3}] emission, and its relatively
high polarization of 7\%, suggest that the scattering region is likely
to be smaller than this estimate by an order of magnitude or more.  If
the polarized [\ion{O}{3}] emission originates from the compact
nuclear source seen in the \hst\ image, then the scattering material
would have to be located within a region of diameter 2.5 pc
surrounding the nucleus.  W95 describe the nuclear polarization as
``marginally resolved'' but do not give a quantitative comparison with
the seeing disk size.  One possible explanation is that the apparent
extension of the nuclear polarization in their image could result from
atmospheric seeing combined with a foreground of interstellar
polarization within NGC 4258, which is visible throughout their map.
Imaging polarimetry in the continuum and the [\ion{O}{3}] line with
\hst\ would provide improved constraints on the size of the scattering
region.  If the torus is aligned with the masing disk and oriented
nearly edge-on, then the scattering material must lie above and/or
below the obscuring torus, rather than within the inner ``hole'' (as
in the model of \cite{hlb97}).

If the scattering region is a cone (or bicone) extending above the
plane of the obscuring torus, then the opening angle of the cone can
be estimated from the degree of polarization of the scattered
radiation.  For a given torus inclination, broad scattering cones
produce lower polarization than narrow cones because the light
reflected from a broad region contains radiation scattered from
different parts of the cone with a wider range of polarization angles.
The degree of polarization of the scattered narrow-line and continuum
emission in NGC 4258 is unknown; the 13.9\% polarization of
[\ion{O}{2}] \lam7325 can be taken as a lower limit to the intrinsic
degree of polarization.  Using as a guide the calculations presented
in Figure 5 of Hines \& Wills (1993), which are based on the method of
Brown \& McLean (1977), for $i$ in the range 80-90\arcdeg\ and $p \geq
14\%$, the scattering cone half-angle is $\theta_c \leq 70\arcdeg$.
If the off-nuclear emission-line regions seen in the \hst\ image are
in fact ionization cones photoionized by the AGN, the widths of these
regions would imply a torus half-opening angle of roughly
$\sim60\arcdeg$ (for the southern region) and $\sim50\arcdeg$ (for the
northern region).

The polarized continuum flux at 5500 \AA\ is $f_\lambda =
1.3\times10^{-17}$ erg cm\persq\ s\per\ \AA\per.  Based on the results
of the starlight subtraction procedure, an upper limit to the
nonstellar continuum flux in total light is $f_\lambda(5500 $\AA$) <
2.1\times10^{-16}$ erg cm\persq\ s\per\ \AA\per.  To be consistent
with this upper limit, the nonstellar continuum must have $p > 6\%$.
A reliable estimate of the intrinsic optical continuum luminosity
would provide valuable constraints on accretion models and on the
accretion rate in NGC 4258 (\cite{gnb99}).  Unfortunately, the unknown
scattering geometry and scattering optical depth render any such
estimates (such as those presented by W95) highly uncertain.  

If there is a thick obscuring torus in NGC 4258, then some fraction of
the UV/X-ray continuum will be intercepted by the torus and
reprocessed to infrared (IR) wavelengths. The far-IR flux of the NGC
4258 nucleus has been measured only at 10 \micron\ (\cite{rl78}), and
part of the observed far-IR emission must be primary radiation from
the accretion flow itself (e.g., \cite{gnb99}), so it is difficult to
place meaningful constraints on the reprocessed IR luminosity.  NGC
1068 is a factor of $\sim200$ brighter than NGC 4258 both in observed
[\ion{O}{3}] flux and at 10 \micron\ (using measurements from
\cite{hfs97}, \cite{so75}, and \cite{rl78}).  However, in NGC 1068 the
[\ion{O}{3}] emission is considered to be unobscured (\cite{mgm91})
while in NGC 4258 the [\ion{O}{3}]-emitting region must be
substantially hidden by the torus.  We caution against drawing any
conclusions based on such a simple comparison, as the properties of
the torus in NGC 4258 are poorly constrained at best and the existing
observations do not allow us to distinguish between the possible
models for the observed IR emission.

W95 found that the polarized continuum shape over 4580--7110 \AA\ was
adequately fit by a $f_\nu \propto \nu^{-1.1}$ power law.  In our
spectra, however, the Stokes flux continuum drops off shortward of
5000 \AA, even after correcting for the amount of polarized flux lost
due to the slit misalignment.  If this turnover is a genuine feature
of the nonstellar continuum, its shape can be used to constrain the
accretion rate (for thin-disk models), or the transition radius
between a thin accretion disk and an advection-dominated accretion
flow (\cite{gnb99}).  Further observations at shorter wavelengths are
needed to determine whether this turnover is indeed a feature of the
nonstellar continuum, or whether it may instead be an artifact
resulting from interstellar polarization within NGC 4258 or the
Galaxy.  The total contribution of interstellar polarization
(including interstellar polarization within NGC 4258) could be
determined, in data of higher signal-to-noise ratio, by measuring the
equivalent widths of stellar absorption features such as \ion{Ca}{2}
H+K or the Mg \emph{b} lines in the Stokes flux spectrum.

\section{Conclusions}

The narrow-line properties of NGC 4258 are unusual in comparison with
those of most Seyfert 2 nuclei in which hidden broad-line regions have
been detected.  Its narrow forbidden lines are highly polarized, with
$p$ as high as 13.9\%, and the degree of polarization is correlated
with the critical density of the transitions.  The
[\ion{O}{3}]-emitting region is compact and centered at the nucleus,
with a nearly unresolved core of FWHM $\lesssim2.5$ pc.

Determining the mechanism responsible for the narrow-line polarization
in NGC 4258 is of relevance to AGN unification models in general, as
it pertains to the question of whether or not \emph{all} Seyfert
nuclei contain a geometrically thick obscuring torus.  Despite the
superficial resemblance between the observed narrow-line polarizations
and the predicted polarization for an edge-on emission source with an
electron-scattering atmosphere (\cite{cha60}), the disk-emission model
probably does not apply to this object; the requirement of high
scattering optical depth is probably unrealistic, and the line
polarizations are too high.  Scattering above the plane of an
obscuring torus, or above a highly warped thin disk, is the most
likely explanation for the narrow-line polarizations in NGC 4258, as
first proposed by W95.  From the compactness of the [\ion{O}{3}]
emission, we estimate that the torus must have a diameter of
$\lesssim2.5$ pc. 

The geometry of the obscuring material relative to the masing disk
remains an open question.  The masing disk may simply be so warped in
its outer regions that it hides the inner portion of the NLR from our
direct view.  If the disk is more nearly flat, then the masing region
could be located at the midplane of the obscuring torus where the
geometry is favorable for maser amplification, or the torus might
surround the masing disk at larger radii.  Alternatively, if the
masing clouds are located in a wind driven from the disk surface, as
in the model of Kartje \etal (1999), then the wind itself would
provide the obscuring material.

There is no direct evidence in our data for a broad component of \hal\
in total or polarized light, but we cannot exclude the possibility
that a broad component may be present.  If the continuum source and
part of the NLR are hidden by an opaque torus, and if NGC 4258 does
contain a BLR, then the BLR should be entirely obscured and the
broad-line emission detected in higher-resolution spectra should
consist entirely of scattered light.  A clear detection of a polarized
broad component of \hal\ would provide strong support for the
obscuring torus interpretation.

\acknowledgements

The W. M. Keck Observatory is operated as a scientific partnership
among the California Institute of Technology, the University of
California, and NASA, and was made possible by the generous financial
support of the W. M. Keck Foundation.  A. J. B. thanks Charles Gammie
and Jules Halpern for helpful discussions.  We are grateful to an
anonymous referee for a careful reading of the manuscript.  Research
by A. J. B. is supported by a postdoctoral fellowship from the
Harvard-Smithsonian Center for Astrophysics.  Research by H.T.,
W.v.B., and M.S.B. at IGPP is performed under the auspices of the US
Department of Energy under contract W-7405-ENG-48.  Research by
A. V. F. is supported by NASA grants NAG 5-3556 and AR-07527.  This
work is based partly on observations with the NASA/ESA Hubble Space
Telescope, obtained from the data archive at the Space Telescope
Science Institute, which is operated by the Association of
Universities for Research in Astronomy, Inc. under NASA contract
No. NAS5-26555.

\clearpage

\begin{center}
Figure Captions
\end{center}

\figcaption[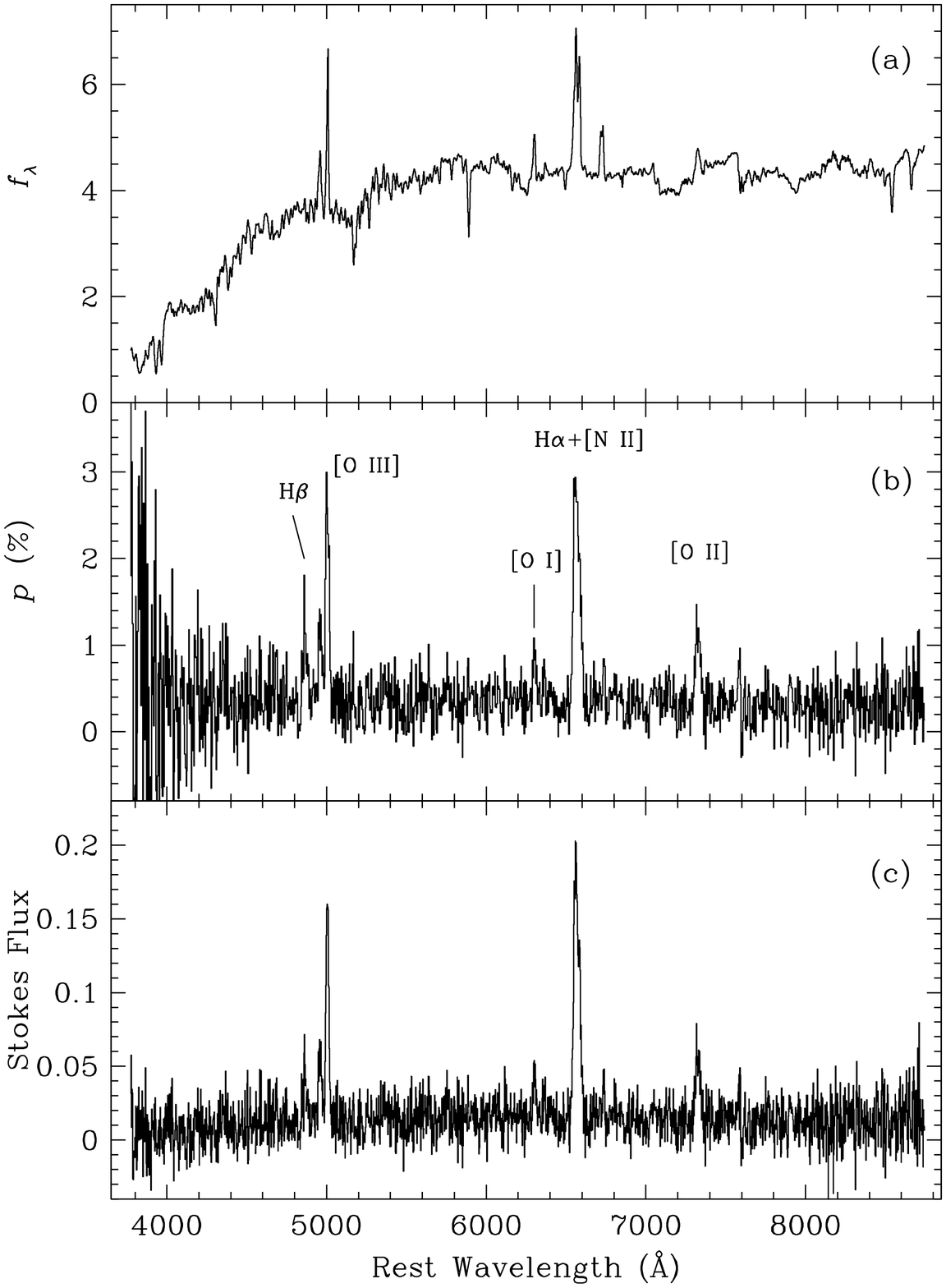]{Polarization data for NGC 4258.  {\it (a)} Total
flux, in units of $10^{-15}$ erg s\per\ cm\persq\ \AA\per.  {\it (b)}
Degree of linear polarization, given as the rotated Stokes
parameter. {\it (c)} Stokes flux, equal to the product of {\it (a)}
and {\it (b)}, in units of $10^{-15}$ erg s\per\ cm\persq\ \AA\per.
For clarity, the spectra have been binned to 4 \AA\
pix\per. \label{4258montage} }

\figcaption[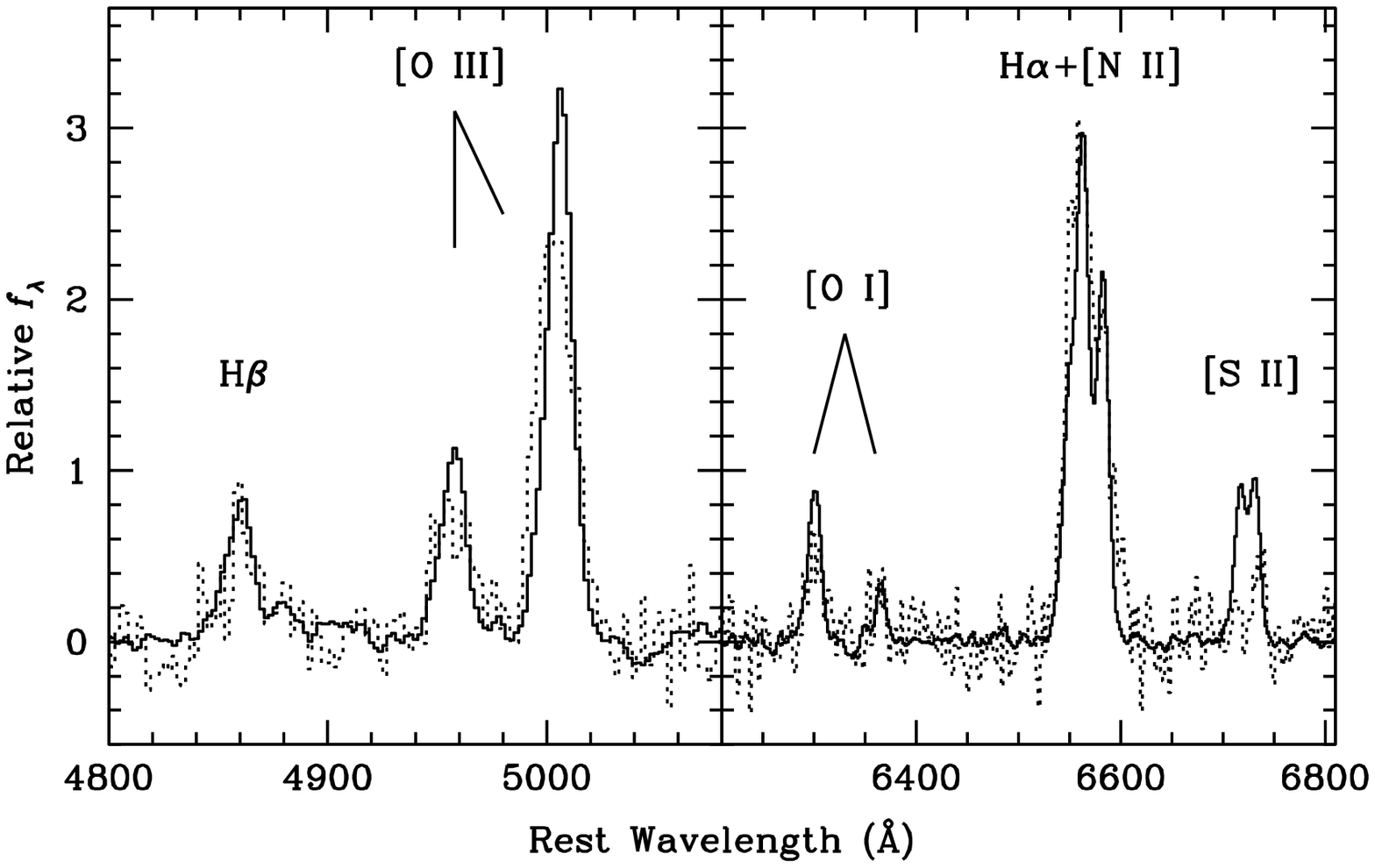]{Comparison of emission-line profiles in total
flux and Stokes flux.  {\it Solid line:} Total flux spectrum after
removal of starlight, in units of $10^{-15}$ erg s\per\ cm\persq\
\AA\per.  {\it Dotted line:} Continuum-subtracted Stokes flux, scaled
by a factor of 16 to achieve a rough match with the emission-line
strengths in total flux for comparison purposes.
\label{4258fluxcomp}}

\figcaption[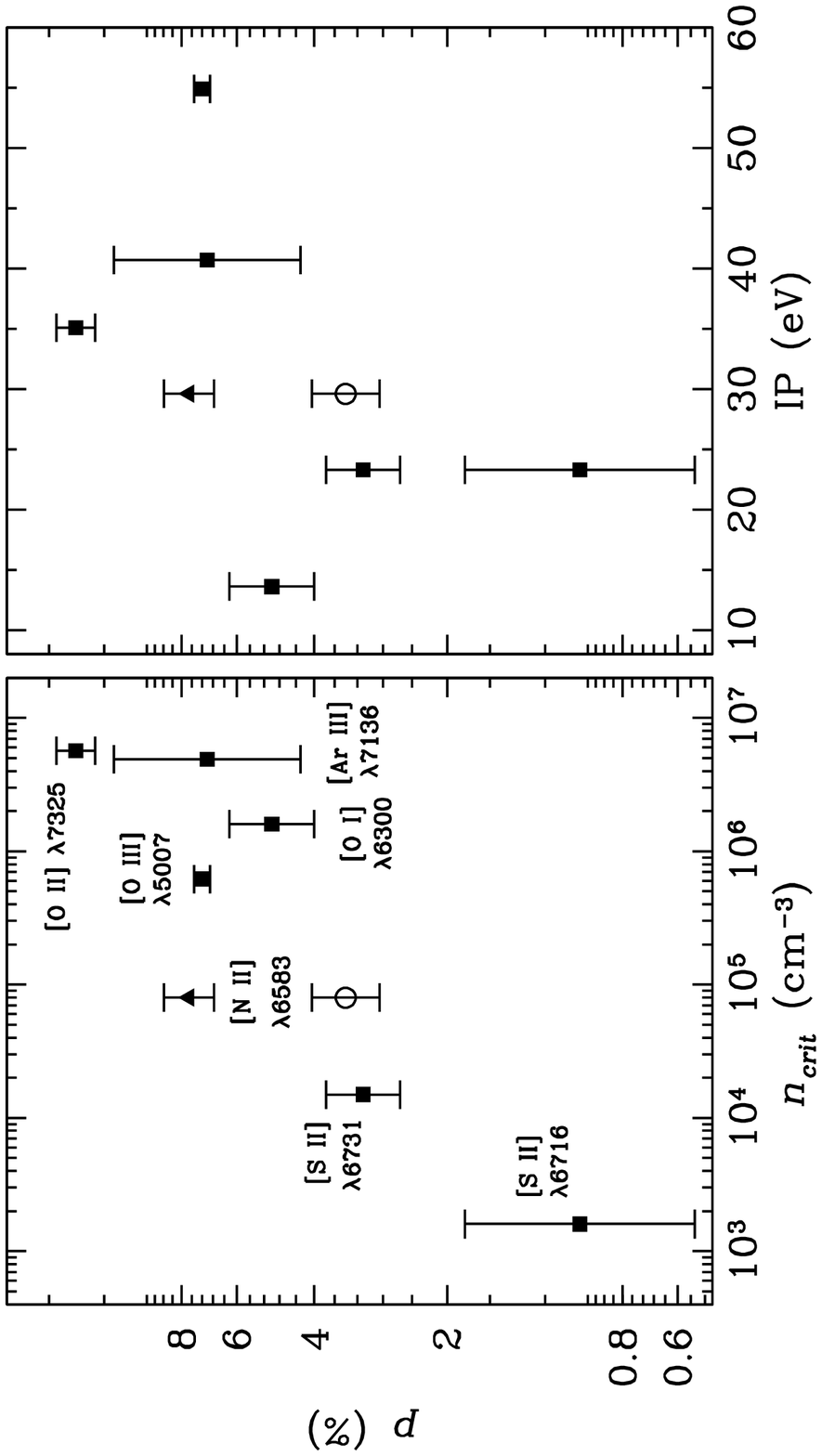]{Forbidden-line polarization as a function of
\ncrit\ and ionization potential.  The open circle represents the
polarization of [\ion{N}{2}] measured from the 4-Gaussian fit
including a broad \hal\ component, while the triangle represents
the [\ion{N}{2}] polarization derived from the 3-Gaussian fit without
a broad \hal\ component.  \label{4258pvsnc}}

\figcaption[fig4.ps]{\emph{Left panel---} Continuum-subtracted \hst\
WFPC2/PC F502N image of NGC 4258 showing emission from [\ion{O}{3}]
$\lambda5007$. \emph{Right panel--- } F547M ($V$-band) PC image of NGC
4258.  The image section is $20\arcsec\times20\arcsec$, and the
rectangle surrounding the nucleus shows the region included in the
$1\arcsec\times3\farcs4$ spectroscopic extraction.  The masing portion
of the disk has a diameter of $\sim0\farcs015$ (Miyoshi \etal 1995),
or about one-third the size of a PC pixel.  At $D=7.3$ Mpc, 1\arcsec\
corresponds to 35 pc.
\label{4258oiii}}

\clearpage
\small

\begin{deluxetable}{lccccccc}
\tablewidth{7in}
\tablenum{1}
\tablecaption{Line and Continuum Properties}
\tablehead{
	\colhead{Line} & 
	\colhead{\ncrit\tablenotemark{a}} & 
	\colhead{IP} & 
	\colhead{$f/f($H$\beta$)\tablenotemark{b}} &
	\colhead{FWHM(tot)\tablenotemark{c}} & 
	\colhead{FWHM(pol)\tablenotemark{d}} & 	
	\colhead{$p$} & 	
	\colhead{$\theta$}  \nl
	\colhead{} & 
	\colhead{(\percucm)} & 
	\colhead{(eV)} & 
	\colhead{} &
	\colhead{(\kms)} & 
	\colhead{(\kms)} & 
	\colhead{(\%)} & 
	\colhead{(\arcdeg)}
}
\startdata
H$\beta$\dotfill                   & \nodata & \nodata & 1.0 & $690 \pm 90$ & $1200 \pm 400$ & $6.3 \pm 1.0$ & $95 \pm 4$ \nl
[\ion{O}{3}] $\lambda4959$\dotfill & $6.2 \times 10^5$ & 54.9 & 1.2 & $520 \pm 60$ & $1240 \pm 220$ & $6.9 \pm 0.7$ & $89 \pm 3$ \nl
[\ion{O}{3}] $\lambda5007$\dotfill & $6.2 \times 10^5$ & 54.9 & 3.3 & $530 \pm 30$ & $1010 \pm 70\phn$ & $7.2 \pm 0.3$ & $89 \pm 1$ \nl
[\ion{O}{1}] $\lambda6300$\dotfill & $1.6 \times 10^6$ & 13.6 & 1.0 & $500 \pm 30$ & $\phn780 \pm 180$ & $5.0 \pm 1.1$ & $91 \pm 7$ \nl
[\ion{O}{1}] $\lambda6363$\dotfill & $1.6 \times 10^6$ & 13.6 & 0.3 & $\phn300 \pm 110$ & \nodata & $5.9 \pm 2.4$ & $\phn76 \pm 11$ \nl
H$\alpha$ + [\ion{N}{2}] blend\dotfill & \nodata & \nodata &  6.7  & \nodata & \nodata & $7.9 \pm 0.2$ & $89 \pm 1$ \nl
H$\alpha$\tablenotemark{e}\dotfill     & \nodata & \nodata & 3.8 & $660 \pm 30$ & $1170 \pm 220$ & $8.0 \pm 0.7$ & $89 \pm 1$ \nl
[\ion{N}{2}] $\lambda6583$\tablenotemark{e}\dotfill & $8.0 \times 10^4$ & 29.6 & 2.0 & $330 \pm 20$ & $1120 \pm 250$ & $7.7 \pm 1.0$ & $90 \pm 2$ \nl
[\ion{S}{2}] $\lambda6716$\dotfill &  $1.6 \times 10^3$ & 23.3 & 0.8 & $310 \pm 20$ & \nodata & $1.0 \pm 0.6$ & $90 \pm 9$ \nl
[\ion{S}{2}] $\lambda6731$\dotfill &  $1.5 \times 10^4$ & 23.3 & 0.9 & $310 \pm 20$ & \nodata & $3.1 \pm 0.6$ & $90 \pm 2$ \nl
[\ion{Ar}{3}] $\lambda7136$\dotfill&  $4.9 \times 10^6$ & 40.7 & 0.2 & $\phn600 \pm 120$ & \nodata & $7.0 \pm 3.4$ & $\phn87 \pm 14$\nl
[\ion{O}{2}] $\lambda7325$\tablenotemark{f}\dotfill & $5.7 \times 10^6$ & 35.1 & 0.8 & $870 \pm 50$ & $1350 \pm 150$ & $13.9 \pm 1.4\phn$ & $89 \pm 3$ \nl
Continuum 4000--4800 \AA\ \dotfill  & & & & & & $0.38 \pm 0.03$ & $74 \pm 2$ \nl
Continuum 5100--6100 \AA\ \dotfill & & & & & & $0.35 \pm 0.01$ & $79 \pm 1$ \nl
Continuum 7500--8500 \AA\ \dotfill & & & & & & $0.29 \pm 0.02$ & $78 \pm 2$
\enddata

\tablenotetext{a}{Critical densities calculated for $T = 10^4$ K using
the IRAF/STSDAS task nebular.ionic.}

\tablenotetext{b}{Total flux of line relative to $f($H$\beta)$ = $1.7
\times 10^{-14}$ erg cm$^{-2}$ s$^{-1}$, not corrected for reddening;
this \hbeta\ flux may include a contribution from the broad-line
component.}

\tablenotetext{c}{Full width at half-maximum of line in total flux,
corrected for instrumental broadening.}

\tablenotetext{d}{Full width at half-maximum of line in Stokes flux,
corrected for instrumental broadening.}

\tablenotetext{e}{For the \hal+[\ion{N}{2}] blend, the component
polarizations listed here are the values derived from the 3-Gaussian
fit which does not include a broad \hal\ component.}

\tablenotetext{f}{[O II] $\lambda7325$ refers to the unresolved
$\lambda\lambda7319,7331$ blend, which may contain some [Ca II]
$\lambda7324$ emission as well.}

\end{deluxetable}

\clearpage

\begin{figure}
\plotone{fig1.ps}
\end{figure}

\begin{figure}
\plotone{fig2.ps}
\end{figure}

\begin{figure}
\plotone{fig3.ps}
\end{figure}

\begin{figure}
\plottwo{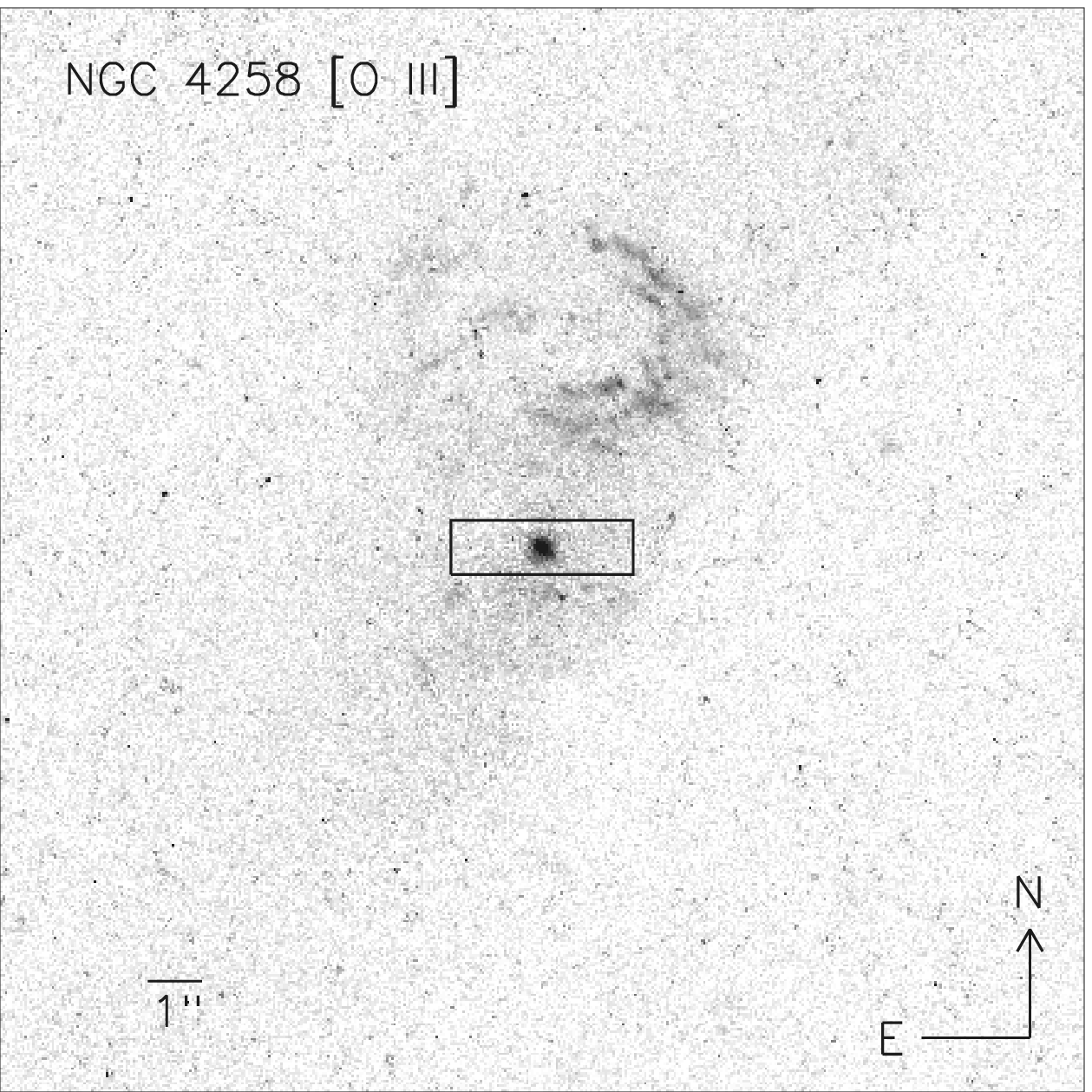}{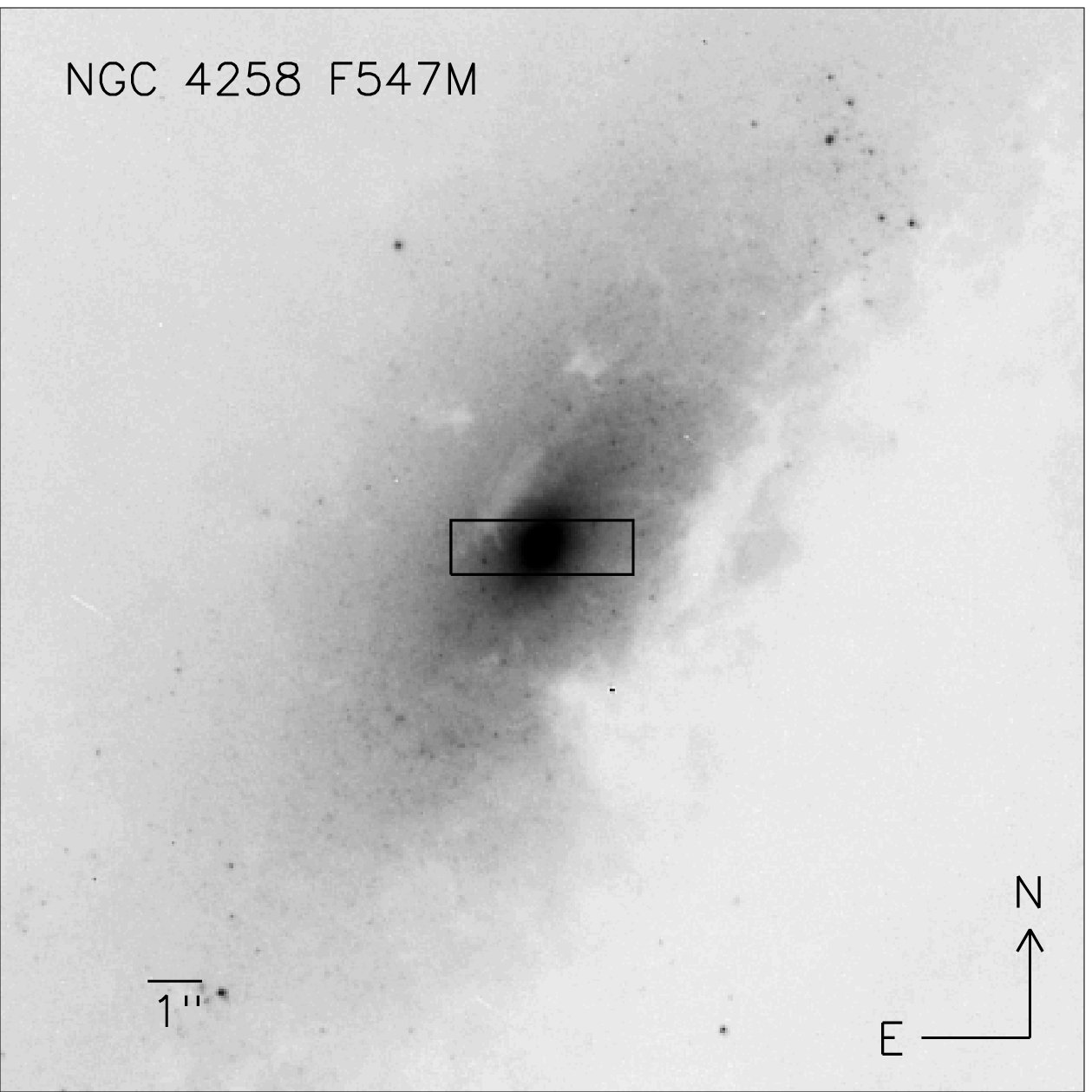}
\end{figure}


\begin{thebibliography}{}

\bibitem[Antonucci \& Miller 1985]{am85} Antonucci, R. R. J., \&
Miller, J. S. 1985, \apj, 297, 621

\bibitem[Brown \& McLean 1977]{bm77} Brown, J. C., \& McLean, I. S. 1977,
\aap, 57, 141

\bibitem[Cecil \etal 1995]{cmv95} Cecil, G., Morse,
J. A., \& Veilleux, S. 1995, \apj, 452, 613

\bibitem[Chandrasekhar 1960]{cha60} Chandrasekhar, S.  1960, Radiative
Transfer (New York: Dover)

\bibitem[Chary \& Becklin 1997]{cb97} Chary, R., \& Becklin, E. E.
1997, \apj, 485, L75

\bibitem[Chen \etal 1997]{cht97} Chen, K., Halpern, J. P., \&
Titarchuk, L. G.  1997, \apj, 483, 194

\bibitem[Clemens \& Tapia 1990]{ct90} Clemens, D. P., \& Tapia, S.
1990, \pasp, 102, 179

\bibitem[Cohen \etal 1997]{coh97} Cohen, M. H., Vermeulen, R. C.,
Ogle, P. M., Tran, H. D., \& Goodrich, R. W.  1997, \apj, 484, 193

\bibitem[Court\`{e}s \& Cruvellier 1961]{cc61} Court\`{e}s, G., \&
Cruvellier, P. 1961, CR Acad. Sci. Paris, 253, 218

\bibitem[Cox \& Downes 1996]{cd96} Cox, P., \& Downes, D. 1996, \apj,
473, 219

\bibitem[Czyzak \etal 1986]{cka86} Czyzak, S. J., Keyes,
C. D., \& Aller, L. H.  1986, \apjs, 61, 159

\bibitem[di Serego Alighieri \etal 1997]{dsa97} di Serego Alighieri,
S., Cimatti, A., Fosbury, R. A. E., \& Hes, R. 1997, \aap, 328, 510

\bibitem[Ferland \& Netzer 1983]{fn83} Ferland, G. J., \& Netzer,
H. 1983, \apj, 264, 105

\bibitem[Filippenko 1982]{fil82} Filippenko, A. V.  1982, \pasp, 94
715

\bibitem[Filippenko 1985]{f85} Filippenko, A. V.  1985, \apj, 289, 475

\bibitem[Filippenko \& Halpern 1984]{fh84} Filippenko, A. V., \&
Halpern, J. P.  1984, \apj, 285, 458

\bibitem[Filippenko \& Sargent 1985]{fs85} Filippenko, A. V., \&
Sargent, W. L. W. 1985, \apjs, 57, 503

\bibitem[Ford \etal 1996]{ftk96} Ford, H., Tsvetanov,
Z., \& Kriss, G.  1996, \baas, 188, 1602

\bibitem[Gammie \etal 1999]{gnb99} Gammie, C. F.,
Narayan, R., \& Blandford, R. 1999, \apj, 516, 177

\bibitem[Goodrich 1992]{goo92} Goodrich, R. W. 1992, \apj, 399, 50

\bibitem[Halpern \& Steiner 1983]{hs83} Halpern, J. P., \& Steiner,
J. E. 1983, \apj, 269, L37

\bibitem[Heisler \etal 1997]{hlb97} Heisler, C. A.,
Lumsden, S. L., \& Bailey, J. A.  1997, \nat, 385, 700

\bibitem[Herrnstein \etal 1996]{hgm96} Herrnstein, J. R., Greenhill,
L. J., \& Moran, J. M.  1996, \apj, 468, L17

\bibitem[Herrnstein \etal 1997a]{her97a} Herrnstein, J. R., Moran,
J. M., Greenhill, L. J., Diamond, P. J., Miyoshi, M., Nakai, N., \&
Inoue, M.  1997a, \apj, 475, L17

\bibitem[Herrnstein \etal 1997b]{her97b} Herrnstein, J., Moran, J.,
Greenhill, L., Inoue, M., Nakai, N., Miyoshi, M., \& Diamond, P.
1997b, \baas, 191, 2507

\bibitem[Hines \& Wills 1993]{hw93} Hines, D. C., \& Wills, B. J.
1993, \apj, 415, 82

\bibitem[Ho \etal 1993]{hfs93} Ho, L. C.,
Filippenko, A. V., \& Sargent, W. L. W.  1993, \apj, 417, 63

\bibitem[Ho \etal 1997a]{hfsp97} Ho, L. C., Filippenko, A. V., Sargent,
W. L. W., \& Peng, C. Y.  1997a, \apjs, 112, 391

\bibitem[Ho \etal 1997b]{hfs97} Ho, L. C., Filippenko, A. V., \&
Sargent, W. L. W. 1997, \apjs, 112, 315

\bibitem[Kartje \etal 1999]{kke99} Kartje, J. F.,
K\"onigl, A., \& Elitzur, M.  1999, \apj, 513, 180

\bibitem[K\"onigl \& Kartje 1994]{kk94} K\"onigl, A., \& Kartje,
J. F.  1994, \apj, 434, 446

\bibitem[Krist \& Hook 1997]{kh97} Krist, J., \& Hook, R.  1997, The
Tiny Tim User's Guide Version 4.4 (Baltimore: STScI)

\bibitem[Makishima \etal 1994]{mak94} Makishima, K., \etal  1994,
\pasj, 46, L77

\bibitem[Malkan \etal 1998]{mgt98} Malkan, M. A., Gorjian,
V., \& Tam, R. 1998, \apjs, 117, 25

\bibitem[Maoz 1998]{maoz98} Maoz, E. 1998, \apj, 494, L181

\bibitem[Miller \etal 1988]{mrg88} Miller, J. S., Robinson, L. B., \&
Goodrich, R. W.  1988, in Instrumentation for Ground-Based Astronomy,
ed. L. B. Robinson (New York: Springer-Verlag), 157

\bibitem[Miller \etal 1991]{mgm91} Miller, J. S., Goodrich, R. W., \&
Mathews, W. G.  1991, \apj, 378, 47

\bibitem[Miyoshi \etal 1995]{miy95} Miyoshi, M., Moran, J.,
Herrnstein, J., Greenhill, L., Nakai, N., Diamond, P,. \& Inoue, M.
1995, \nat, 373, 127

\bibitem[Nelson \etal 1996]{nel96} Nelson, C. H., MacKenty, J. W.,
Simkin, S. M., \& Griffiths, R. E.  1996, \apj, 466, 713

\bibitem[Neufeld \& Maloney 1995]{nm95} Neufeld, D. A., \& Maloney,
P. R.  1995, \apj, 447, L17

\bibitem[Oke \etal 1995]{oke95} Oke, J. B., \etal  1995, \pasp, 107,
375

\bibitem[Pelat \etal 1981]{paf81} Pelat, D., Alloin, D.,
\& Fosbury, R. A. E.  1981, \mnras, 195, 787

\bibitem[Rieke \& Lebofsky 1978]{rl78} Rieke, G. H., \& Lebofsky,
M. J.  1978, \apj, 220, L37

\bibitem[Schlegel \etal 1998]{sfd98} Schlegel, D. J.,
Finkbeiner, D. P., \& Davis, M.  1998, \apj, 500, 525

\bibitem[Serkowski \etal 1975]{smf75} Serkowski, K., Mathewson, D. S.,
\& Ford, V. L.  1975, \apj, 196, 261

\bibitem[Shields \& Oke 1975]{so75} Shields, G. A., \& Oke,
J. B. 1975, \apj, 197, 5

\bibitem[Shuder 1981]{shu81} Shuder, J. M.  1981, \apj, 244, 12

\bibitem[Terndrup \etal 1984]{tls84} Terndrup, D., Lauer,
T. R., \& Stover, R. 1984, Lick Obs. Tech. Rep. No. 33

\bibitem[Tran 1995]{tr95} Tran, H. D. 1995, \apj, 440, 565

\bibitem[Wilkes \etal 1995]{wil95} Wilkes, B. J., Schmidt, G. D.,
Smith, P. S., Mathur, S., \& McLeod, K. K.  1995, \apj, 455, L13 (W95)

\bibitem[Young \etal 1996]{you96} Young, S., Hough, J. H., Efstathiou,
A., Wills, B. J., Bailey, J. A., Ward, M. J., \& Axon, D. J.  1997,
\mnras, 281, 1206

\end{thebibliography}
\end{document}